\documentclass[prb,twocolumn,amsmath,amssymb,showpacs]{revtex4}

\usepackage{graphicx}
\usepackage{dcolumn}
\usepackage{bm}
\usepackage{mathrsfs}
\usepackage{appendix}
\usepackage{bbm}
\usepackage{color}

\def \vS {{\bf S}}
\def \vR {{\bf R}}
\def \ve {{\bf e}}
\def \vQ {{\bf Q}}
\def \vq {{\bf q}}

\def \ve {{\bf e}}
\def \vm {{\bf m}}

\def \mb {\mu_{\rm B}}

\def \xp {{\bf x}^{\prime }}
\def \yp {{\bf y}^{\prime }}
\def \zp {{\bf z}^{\prime }}
\def \xx {{\bf x}}
\def \yy {{\bf y}}
\def \zz {{\bf z}}

\def \vP {{\bf P}}
\def \BF {{\rm BiFeO$_3$} }
\def \BP {{\rm BiFeO$_3$}}

\def \xq {x^{\prime}}
\def \yq {y^{\prime}}
\def \zq {z^{\prime}}

\def \vB {{\bf B}}

\def \hi {{h_i}}

\def \XX {{\bf X}}
\def \YY {{\bf Y}}
\def \ZZ {{\bf Z}}

\begin{document}

\title{The Microscopic Model of BiFeO$_3$}

\author{R.S. Fishman}

\affiliation{Materials Science and Technology Division, Oak Ridge National Laboratory, Oak Ridge, Tennessee 37831, USA}

\date{\today}

\begin{abstract}

Many years and great effort have been spent constructing the microscopic model for the room temperature multiferroic \BP.
However, earlier models implicitly assumed that the cycloidal wavevector $\vq $ was confined to one of the three-fold symmetric axis in the hexagonal plane normal to the electric
polarization.  Because recent measurements indicate that $\vq $ can be rotated by a magnetic field, it is essential to properly treat the anisotropy that 
confines $\vq $ at low fields.  We show that the anisotropy energy $-K_3 S^6 \sin^6\theta \,\cos 6 \phi $ confines the wavevectors $\vq $ to the three-fold axis
$\phi =0$ and $\pm 2\pi/3$ within the hexagonal plane with $\theta = \pi /2$.

\end{abstract}


\maketitle

Multiferroics have attracted a great deal of attention due to their possible technological applications.  In multiferroic materials, the 
magnetization can be controlled by an electric field and the electric polarization can be controlled by a magnetic field.
The ability to reverse the voltage with a magnetic field 
offers the possibility of magnetic storage without Joule heating loss due to electrical currents [\onlinecite{eer06, zhao06}].  
To take advantage of this capability, however, we must first learn how to manipulate magnetic domains with a magnetic field.  

In type I multiferroics, magnetic order develops at a lower temperature than the ferroelectric
polarization.  In type II multiferroics, the electric polarization directly couples to the magnetic state [\onlinecite{khomskii06}] and the two
develop at the same temperature.  The coupling between electrical and magnetic properties 
is typically stronger in type II multiferroics but type I multiferroics have much higher transition temperatures.
To date, the highest magnetic transition temperature has been found in the type I multiferroic BiFeO3 with $T_N \approx Å 640 K$ [\onlinecite{sosnowska82}].

\begin{figure}
\includegraphics[width=8cm]{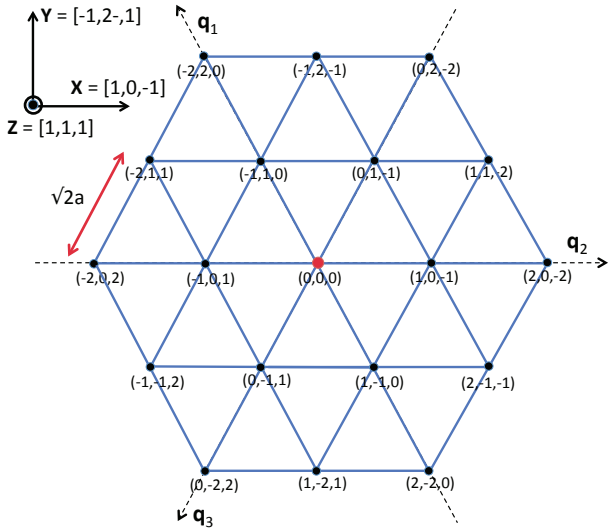}
\caption{(Color online) A hexagonal plane normal to $\ZZ $.  In zero field, three domains with
wavevectors $\vq_k$ are stable.  Points $(n_1,n_2,n_3)$ label sites $\vR /a = n_1 \xx + n_2 \yy +n_3 \zz $.}
\end{figure}

The long-wavelength spin cycloid of \BF [\onlinecite{sosnowska82, herrero10}] has wavector $\vQ + \vq $ where $\vQ = (\pi /a)(1,1,1)$ 
is the antiferromagnetic reciprocal lattice vector in terms of the lattice constant $a \approx 3.96$ \AA $\,$ of the 
pseudo-cubic unit cell.
If $\vq =0 $, then the spin state of \BF would be a G-type antiferromagnet.  The wavelength of the spin cycloid 
$\lambda = 2\pi /q$ is about 62 nm and its spins lie primarily in the plane defined by the electric polarization $\vP $ and the wavevector
$\vq $.  There are three possible magnetic domains with $\vq $ lying along one of the three-fold symmetric axis normal to $\vP $, which itself lies along 
one of the cubic diagonals.  With $\vP \parallel \zp = [1,1,1]$ ($[a,b,c]$ is a unit vector normalized to 1), the 
wavevectors $\vq $ can lie along the $[-1,1,0]$, $[1,0,-1]$, or $[0,-1,1]$ directions in zero field.  The hexagonal plane
normal to $[1,1,1]$ is sketched in Fig.1, with points given by $\vR  = a(n_1 \xx + n_2 \yy +n_3 \zz )$ 
in terms of the integers $n_i$.  All points in this hexagonal satisfy $\zp \cdot \vR =0$ or $n_1+n_2+n_3=0$.

Previous microscopic models for \BF such as used in Ref.[\onlinecite{fishman13c}]  implicitly 
assumed that the domain wavevector $\vq $ remains fixed along one of the three-fold axis in a magnetic field.
Because the magnetic susceptibility perpendicular to $\vq $ is much
larger than the susceptibility parallel to $\vq $ [\onlinecite{leb07}],
a magnetic field $\vB $ favors domains with $\vq \perp \vB $.  
Recent evidence [{\onlinecite{bordacsun}] reveals that a magnetic 
field rotates the wavevectors $\vq $ within the hexagonal plane away from the three-fold axis towards an orientation 
perpendicular to $\vB $.

Our recently revised Hamiltonian is valid for any $\vq $ and given by 
\begin{eqnarray}
\label{Ham}
&&{\cal H} = -J_1\sum_{\langle i,j\rangle }\vS_i\cdot \vS_j -J_2\sum_{\langle i,j \rangle'} \vS_i\cdot \vS_j
\nonumber \\
&&+D_1\, \sum_{\langle i,j\rangle } (\zp \times {\bf e}_{i,j}/a) \cdot (\vS_i\times\vS_j) \nonumber \\
&& + D_2 \, \sum_{\langle i,j\rangle } \, (-1)^\hi \,\zp \cdot  (\vS_i\times\vS_j)\nonumber \\
&& -K_1\sum_i (\zp \cdot \vS_i )^2
- 2\mb B \sum_i \vm \cdot \vS_i ,
\end{eqnarray}
where $\ve_{i,j } = a{\bf x}$, $a{\bf y}$, or $a{\bf z}$ connects the $S=5/2$ spin $\vS_i$ on site $\vR_i$ with nearest-neighbor spins $\vS_j$ 
on site $\vR_j=\vR_i + \ve_{i,j}$.  The integer $h_i=\sqrt{3} \vR_i \cdot \zp /a$ is the hexagonal layer number.   
Antiferromagnetic exchange interactions $J_1$
and $J_2$ were determined from inelastic neutron-scattering measurements [\onlinecite{jeong12, matsuda12, xu12}].
The $K_1$ anisotropy term provides an easy axis along $\zp $ and can be estimated from the intensity of the third cycloidal harmonic [\onlinecite{zal00}].

Two Dzyalloshinskii-Moriya interactions are produced by broken inversion symmetry.
While the first DM interaction $D_1$ determines the cycloidal period $\lambda $ [\onlinecite{sosnowska95}],
the second DM interaction $D_2$ creates the small tilt $\tau $ of the cycloid out of the plane defined by $\zp $ and $\vq $
[\onlinecite{sosnowska95, pyat09}].  Because this tilt averages to zero over the cycloid, \BF has no net ferrimagnetic moment below 
about 18 T.  Above 18 T,  \BF undergoes a transition into a canted G-type antiferromagnet [\onlinecite{tokunaga10}] 
with a small ferrimagnetic moment perpendicular to $\vP $.  
Unlike in earlier models [\onlinecite{fishman13c}], the DM terms only involve sums over nearest neighbors.
For convenience, we summarize all of these energies, their values [\onlinecite{param}], 
and the experimental or theoretical methods used for their
determination in Table II.

\begin{table}
\caption{Reference frames of BiFeO$_3$}
\begin{ruledtabular}
\begin{tabular}{cc}
unit vectors & description \\
\hline
$\{\xx, \yy  ,\zz \}$ & pseudo-cubic unit vectors  \\
& $\xx = [1,0,0]$, $\yy = [0,1,0]$, $\zz = [0,0,1]$ \\
$\{\xp ,\yp ,\zp \}$ & rotating reference frame of cycloid \\
&  $\xp \parallel \vq $, $\zp = [1,1,1]$, $\yp = \zp \times \xp $ \\
$\{ \XX , \YY , \ZZ \}$ & fixed reference frame of hexagonal plane \\
& $\XX = [1,0,-1]$, $\YY = [-1,2,-1]$, $\ZZ = [1,1,1]$\\
\end{tabular}
\end{ruledtabular}
\end{table}

\begin{table*}
\caption{Exchange and anisotropy parameters of \BF [\onlinecite{param}]}
\begin{ruledtabular}
\begin{tabular}{ccccc}
parameter & description & value & order in $l$ &method for determination \\
\hline
$J_1$ & nearest-neighbor exchange & -5.3 meV& 0 & inelastic neutron scattering [\onlinecite{jeong12, matsuda12, xu12}] \\
$J_2$ & next nearest-neighbor exchange & -0.2 meV & 0 & inelastic neutron scattering [\onlinecite{jeong12, matsuda12, xu12}] \\
$D_1$ & first DM interaction & 0.18 meV& 1 & cycloidal wavelength [\onlinecite{sosnowska95}] \\
$D_2$ & second DM interaction & 0.06 meV & 1 & cycloidal tilt [\onlinecite{tokunaga10, rama11}], spin-wave modes [\onlinecite{ruette04, fishman13a, jeong14}]\\
$K_1$ & easy-axis anisotropy & 0.004 meV $\,$ & 2 & third cycloidal harmonic [\onlinecite{zal00}], high-field diffraction [\onlinecite{ohoyama11}], \\
 & & & & spin-wave modes [\onlinecite{matsuda12, nagel13, fishman13a, jeong14}], tight binding [\onlinecite{desousa13}] \\
 $K_3$ & three-fold anisotropy & $\sim 6 \times 10^{-6}$ meV $\,\,\,\,\,\,\,\,\,\,\,\,\,\,\,$ & 4 & domain rotation in a magnetic field \\
\end{tabular}
\end{ruledtabular}
\label{all}
\end{table*}

The model in Eq.(\ref{Ham}) was constructed so that it reduces to previous models when $\vq $ lies along any of the three-fold axis.
But there is a problem.  Because the revised model is rotationally invariant, $\vq $ can point along any direction in the hexagonal
plane normal to $\zp $ with no cost in energy!  This can be easily seen from Eq.(\ref{Ham}), which involves the polarization direction $\zp $ but not
the two orthogonal vectors $\xp \parallel \vq $ and $\yp = \zp \times \xp $.

In the local reference frame defined by $\xp $, $\yp $, and $\zp$,  a spin cycloid along any wavevector $\vq$ can be approximated by
\begin{eqnarray}
\label{syc1}
S_{\xq }(\vR_i)&=& S (-1)^{h_i+1} \cos \tau \sin \big(\vq \cdot\vR_i\big), \\
\label{syc2}
S_{\yq }(\vR_i)&=& S \sin \tau \sin \big(\vq \cdot \vR_i\big), \\
\label{syc3}
S_{\zq }(\vR_i)&=&S (-1)^{h_i+1} \cos \big( \vq \cdot \vR_i\big).
\end{eqnarray}
While susceptibility measurements [\onlinecite{tokunaga10}] indicate that
$\tau \sim 0.3^{\circ }$, neutron-scattering measurements [\onlinecite{rama11}] indicate that $\tau \sim 1^{\circ }$ is
about three times larger.

Although they ignore higher harmonics $(2m+1)\vq \cdot \vR_i $ ($m > 1$) produced by the easy-axis anisotropy and the 
second DM interaction, these simplified expressions are useful for taking averages over the lattice.   
The error introduced by neglecting higher harmonics is of order $C_3/C_1 \approx 5\times 10^{-3} $ where $C_{2m+1}$ are the 
coefficients for the $2m+1$ harmonic [\onlinecite{fishman13a}].   Only odd harmonics contribute in zero field and those harmonics fall off rapidly with 
$2m+1$.

To avoid confusion with the $\{ \xp ,\yp ,\zp \}$ reference frame of the cycloid, 
we define $\XX = [1,0,-1]$ and $\YY = [-1,2,-1]$ as fixed axis in the hexagonal
plane normal to $\ZZ = \XX \times \YY = [1,1,1]$.  Of course, $\ZZ = \zp $ lies along $\vP $.   The different reference frames for
\BF are summarized in Table I.

To lift the rotational invariance of the microscopic model constructed above, we consider all possible anisotropy 
terms consistent with the $R$3$c$ rhombohedral symmetry of \BF [\onlinecite{weingart12}].  Up to order $S^6$, those terms are
\begin{eqnarray}
{\cal H}_{K_1}&=& - K_1 \sum_i {S_{iZ}}^2,\\
{\cal H}_{K_2}&=& -\frac{1}{2} K_2 \sum_i S_{iZ} \Bigl\{ \bigl(S_{iX}+iS_{iY}\bigr)^3 \nonumber \\
&&+  \bigl(S_{iX}-iS_{iY}\bigr)^3\Bigr\}, \\
{\cal H}_{\bar{K}_2}&=& - \bar{K}_2 \sum_i {S_{iZ}}^4,\\
{\cal H}_{K_3}&=& -\frac{1}{2} K_3 \sum_i \Bigl\{ \bigl(S_{iX}+iS_{iY}\bigr)^6 \nonumber \\ 
&&+ \bigl(S_{iX}-iS_{iY}\bigr)^6\Bigr\}, \\
{\cal H}_{\bar{K}_3}&=& - \bar{K}_3 \sum_i {S_{iZ}}^6.
\end{eqnarray}
In terms of the spin-orbit coupling constant $l\vert J_1\vert $ where $l \ll1$, the DM interactions $D_1$ and $D_2$ are of order $l\,\vert J_1\vert $
and the anisotropy constants $K_m$ and $\bar{K}_m$ are of order $l^{m+1}\,\vert J_1\vert $ [\onlinecite{bruno89}].

The anisotropy terms have classical energies $E_K = \langle {\cal H}_K \rangle $:
\begin{eqnarray}
E_{K_1}&=& -S^2 K_1 \sum_i \cos^2 \theta_i, \\
E_{K_2}&=& -S^4 K_2 \sum_i \cos\theta_i \,\sin^3 \theta_i \,\cos 3\phi_i,\\
E_{\bar{K}_2}&=&  -S^4 \bar{K}_2 \sum_i \cos^4 \theta_i, \\
E_{K_3}&=& -S^6 K_3 \sum_i \sin^6 \theta_i \,\cos 6\phi_i,\\
E_{\bar{K}_3}&=&-S^6 \bar{K}_3 \sum_i \cos^4 \theta_i ,
\end{eqnarray}
where the angles $\theta_i$ and $\phi_i$ of the spin
\begin{equation}
\langle \vS_i \rangle = S\Bigl\{ \cos\phi_i \sin\theta_i \,\XX + \sin\phi_i \sin \theta_i \,\YY + \cos \theta_i \,\ZZ \Bigr\}
\end{equation}
are defined in the fixed reference frame defined above.
Other anisotropy energies such as $S^2 K_1^{\prime } \sum_i \sin^2 \theta_i \,\cos 2\phi_i $ and $S^4 K_2^{\prime }\sum_i \sin^4 \theta_i \,\cos 4\phi_i $ 
vanish for the $R$3$c$ crystal structure of \BF [\onlinecite{weingart12}].

Like $E_{K_1}$, $E_{\bar K_2}$ and $E_{\bar K_3}$ strengthen or weaken the easy-axis anisotropy along
$\ZZ $.  Because these three energies have qualitatively the same effects and are very hard to disentangle,
we neglect $\bar{K}_2$  and $\bar{K}_3$.  

Using the expressions for the cycloid in Eqs.(\ref{syc1}-\ref{syc3}), 
we find that $E_{K_2}=0$.  Consequently, $K_2$ will distort the cycloid to produce an 
energy reduction of order $(K_2)^2/\vert J_1\vert \sim l^6 \,\vert J_1 \vert $.  So $E_{K_2} \sim l^6 \, \vert J_1\vert $ can be neglected 
compared to $E_{K_3} \sim l^4 \, \vert J_1 \vert $.  A firm estimate for $K_3$ will have to wait until we report results for the 
metastability of cycloidal domains in a magnetic field.  But assuming that $l\approx 0.1$, 
$S^6 K_3 \sim l^2\, S^4 K_1$ or $K_3 \sim 6 \times 10^{-6}$ meV as in Table II.

It may be necessary to slightly modify the estimates in Table II for $K_1$ and $D_2$ to compensate for the effect of $K_3$.  
While $D_2$ favors the spins to lie perpendicular to $\ZZ $, $K_1 > 0$ favors the spins to lie along $\ZZ $.  
Based on fits to the spectroscopic modes [\onlinecite{fishman13a}], the net anisotropy
favors the spins along $\ZZ $.  Regardless of its sign, the new anisotropy $K_3$ favors the spins to lie in the $\XX - \YY $ 
plane rather than along $\ZZ $.  The energy difference between a spin lying along $\ZZ $ and along a three-fold axis like $\XX $ is 
$-S^6 K_3$.  So to offset the effect of $K_3$, $K_1$ must be increased by $S^4 K_3$.  
For $K_1 \approx 0.004$ meV and $K_3 \approx 6 \times 10^{-6}$ meV, $\Delta K_1 \approx 2.3\times 10^{-4}$ meV 
constitutes an increase of about 6\%.

How does this estimate for $K_3$ in \BF compare with that in other materials?  
The three-fold anisotropy constant $K_3$ can be estimated from the angular dependence of the basal-plane magnetization or the torque.   
For Co$_2$$Y$ ($Y$ = Ba$_2$Fe$_{12}$O$_{22}$) and Co$_2$$Z$ ($Z$ = Ba$_3$Fe$_{24}$O$_{41}$), $\tilde K_3 \equiv S^6 K_3 /V_c \approx 600$ erg/cm$^3$ and 
$1500$ erg/cm$^3$, respectively [\onlinecite{bickford60}] ($V_c$ is the unit cell volume with one magnetic ion).   
Torque measurements were used to estimate [\onlinecite{paige84}] that $\tilde K_3 \approx 1.2 \times 10^5$ erg/cm$^3$ for pure Co.
Anisotropy energies are much larger for rare earths than for transition-metal oxides [\onlinecite{rhyne72}].
While $\tilde K_3 \approx 6300$ erg/cm$^3$ for Gd, it is about 1000 times higher for the heavier rare earths Tb, Dy, Ho, Er, and Tm.
An anisotropy of $K_3 = 6 \times 10^{-6}$ meV for \BF corresponds to $\tilde K_3 = 4\times 10^4$ erg/cm$^3$, larger than for Gd but smaller than for pure Co
or the heavy rare earths.

To conclude, we have added an additional anisotropy energy to the ``canonical" model for \BF in order to lift its rotational invariance in the hexagonal plane
normal to the polarization.  While the anisotropy constant is quite small, it is comparable to that measured in other materials.  Future work will demonstrate
that this three-fold anisotropy has a profound effect on the rotation of domains in a magnetic field.

Thanks to Istvan K\'ezsm\'arki for helpful discussions.
Research sponsored by the U.S. Department of Energy, 
Office of Basic Energy Sciences, Materials Sciences and Engineering Division.


\begin{thebibliography}{}
\bibitem{param} These parameters reproduce the zero-field mode frequencies scaled by $S=5/2$ rather than by $\sqrt{S(S+1)}\approx 3$.
So they may be about 15\% lower than earlier estimates.
\end{thebibliography}
\end{document}